\documentclass[prl,twocolumn,showpacs,preprintnumbers,amsmath,amssymb,superscriptaddress]{revtex4-1}

\usepackage{amsmath,amssymb,amsfonts,mathrsfs} 	
\usepackage{graphicx}
\usepackage{subfigure}
\usepackage{xcolor}
\usepackage[normalem]{ulem}
\usepackage[colorlinks=true,linkcolor=blue,citecolor=blue,urlcolor=black]{hyperref}

\newcommand{\ei}[1]{{\rm e}^{i #1}}
\newcommand{\emi}[1]{{\rm e}^{-i #1}}
\newcommand{\kk}{\mbox{\boldmath$\kappa$}}

\def\beq{\begin{equation}}
\def\eeq{\end{equation}}
\def\bea{\begin{eqnarray}}
\def\eea{\end{eqnarray}}
\def\nn{\nonumber\\}

\def\ket#1{\vert#1\rangle}

\def\me#1#2#3{\langle#1\vert#2\vert#3\rangle}
\def\ev#1{\langle#1\rangle}

\def\EE{{\cal E}}

\def\EEE{\mbox{\boldmath${\cal E}$}}
\def\r{{\bf r}}
\def\v{{\bf v}}
\def\k{{\bf k}}

\def\p{{\bf p}}

\def\P{{\bf P}}

\def\P0{ \Psi_0(0)}

\renewcommand{\[}{\begin{equation}}
\renewcommand{\]}{\end{equation}}

\newcommand{\equ}[1]{Eq.~(\ref{#1})}
\newcommand{\eqs}[2]{Eqs.~(\ref{#1}) and (\ref{#2})}

\def\runtime{(\the\time)\qquad\the\month/\the\day/\the\year}
\def\today
 {\count10=\year\advance\count10 by -2000 \number\day--\ifcase
  \month \or Jan\or Feb\or Mar\or Apr\or May\or Jun\or
             Jul\or Aug\or Sep\or Oct\or Nov\or Dec\fi--\number\count10}

\def\hour{\count10=\time\count11=\count10
\divide\count10 by 60 \count12=\count10
\multiply\count12 by 60 \advance\count11 by -\count12\count12=0
\number\count10 :\ifnum\count11 < 10 \number\count12\fi\number\count11}
\def\draft1#1{
\setlength{\unitlength}{1cm}
\noindent \begin{picture}(0,0)
\put(0,2.5){\noindent \sf DRAFT: {\bf #1} run through \LaTeX\ on \today\ at \hour}
\end{picture}
}

\begin{document}

\title{Linear and nonlinear Hall conductivity in presence of interaction and disorder}
\author{Raffaele Resta}
\email{resta@iom.cnr.it.it}
\affiliation{Istituto Officina dei Materiali IOM-CNR, Strada Costiera 11, 314151 Trieste, Italy}
\affiliation{Donostia International Physics Center, 20018 San Sebasti{\'a}n, Spain}

\date{\today}

\begin{abstract}
The theory of the nonlinear Hall effect has been established by I. Sodemann and L. Fu [Phys. Rev. Lett. 115, 216806 (2015)] in a semiclassical framework: therein, the effect appears as a geometrical property of Bloch electrons, originating from their anomalous velocity. Here I present a more general theory, addressing correlated and/or noncrystalline systems as well, where the expressions of both linear and nonlinear Hall conductivities originate from the many-electron anomalous velocity. The independent-electron results are retrieved as special cases.
\end{abstract}

\maketitle

It is known since long time that transverse dc conductivity is allowed---to linear order in the field---only in materials which spontaneously break time-reversal (T) symmetry: it goes then under the name of anomalous Hall conductivity (AHC) \cite{Nagaosa10}. More recently it has been pointed out that second-order transverse dc conductivity can be nonzero even in T-symmetric materials, provided that inversion (I) symmetry is absent: the quadratic dc response is then called nonlinear Hall conductivity (NHC); the theory so far is based on geometrical concepts at the independent-electron level for crystalline systems, and the relevant expressions are obtained semiclassically \cite{Sodemann15,Matsyshyn19,Nandy19}. In this Letter I show how to formulate the theory at a much more general level, encompassing correlated and/or disordered systems as well. Even in the present case the theory is based on geometrical concepts, although in a many-body framework: in particular on the many-body Berry curvature, whose root is in a seminal paper by Niu and Thouless \cite{Niu84}. The known independent-electron NHC formula \cite{Sodemann15,Matsyshyn19}  will be retrieved as a special case; a few other known results will be also presented en passant, obtained here via somewhat unconventional proofs.

The independent-electron geometrical theory for a pristine crystal only provides the intrinsic AHC term; extrinsic terms are necessarily present in the case of metals \cite{Nagaosa10}. The present formulation allows in principle for the inclusion of disorder and accounts therefore for a part of the extrinsic effects as well, thus generalizing a previous work at the independent-electron level \cite{rap149}.

An outstanding qualitative difference exists between AHC and NHC. In the former case the geometrical intrinsic term is nondissipative: it yields a dc current {\it without} any mechanism accounting for dissipation (e.g. relaxation times); in the latter case, instead, the geometrical expressions yield a transverse {\it free acceleration}; one gets a dc current only after some dissipation mechanism is accounted for. NHC can therefore be assimilated to a skewed nonlinear Drude-like conductivity.

The starting point of the present theory is a milestone paper published by Kohn in 1964 \cite{Kohn64}. Following him, we consider a system of $N$ interacting $d$-dimensional electrons in a cubic box of volume $L^d$, and the family of many-body Hamiltonians parametrized by $\kk$, called ``flux'' or ``twist'': \[ \hat{H}_{\kk} = \frac{1}{2m} \sum_{i=1}^N \left[\p_i + \frac{e}{c} {\bf A}^{(\rm micro)}(\r_i) + \hbar \kk \right]^2 + \hat{V}, \label{kohn} \] where $\hat{V}$ includes the one-body potential (possibly disordered) and electron-electron interaction, while the microscopic vector potential ${\bf A}^{(\rm micro)}(\r)$ summarizes all the intrinsic T-breaking terms, as e.g. those due to spin-orbit coupling to a background of local moments. 
We assume the system to be macroscopically homogeneous; the eigenstates $\ket{\Psi_{n\kk}}$ are normalized to one in the hypercube of volume $L^{Nd}$. 
The thermodynamic limit $N \rightarrow \infty$, $L \rightarrow \infty$, $N/L^d=n$ constant, is understood throughout this Letter. In order to simplify notations we will set $\hat{H}_{0} \equiv \hat{H}$, $\ket{\Psi_{n0}} \equiv \ket{\Psi_{n}}$ , $E_{n0} \equiv E_n$.

We assume Born-von-K\`arm\`an periodic boundary conditions (PBCs): the many-body wavefunctions are periodic with period $L$ over each electron coordinate $\r_i$ independently; the
potential $\hat{V}$ and the intrinsic vector potential ${\bf A}^{(\rm micro)}(\r)$ enjoy the same periodicity. The flux $\kk$---cast into inverse-length dimensions for convenience---corresponds to perturbing the Hamiltonian with a vector potential ${\bf A} = \hbar c \kk /e$, constant in space. Kohn only considered a time-independent $\kk$, which amounts to a pure gauge transformation; the latter has nontrivial effects, given that PBCs violate gauge-invariance in the conventional sense \cite{Kohn64}.
Here we additionally consider even a time-dependent flux, which amounts to perturbing the Hamiltonian with the macroscopic field $\EEE(t) = - \dot{\bf A}(t)/c = - \hbar \dot\kk(t)/e$.

The kinetic-energy term in \equ{kohn} defines the extensive many-electron velocity as \[ \hat{\v}_{\kk} = \frac{1}{m} \sum_{i=1}^N \left[\p_i + \frac{e}{c} {\bf A}^{\rm (micro)}(\r_i) + \hbar \kk\right] = \frac{1}{\hbar} \partial_{\kk} \hat{H}_{\kk} . 
\] When $\kk$ is adiabatically varied in time the instantaneous current density is the sum of two terms: the expectation value of the current operator, and the Niu-Thouless adiabatic current \cite{Niu84,Xiao10}. Their expression is cast as:
\bea  j_\alpha &=& - \frac{e}{ \hbar L^d} \me{\Psi_{0\kk}}{\partial_{\kappa_\alpha} \hat{H}_{\kk}}{\Psi_{0\kk}} \nn &+& \frac{ie}{L^d}( \ev{\partial_{\kappa_\alpha}{\Psi}_{0\kk} |  \dot{\Psi}_{0\kk} } - \ev{\dot{\Psi}_{0\kk} | \partial_{\kappa_\alpha} \Psi_{0\kk} } ) \nn  &=& - \frac{e}{L^d} \left (\frac{1}{\hbar}\partial_{\kappa_\alpha} E_{0\kk} - \Omega_{\alpha\beta}(\kk) \dot{\kappa}_\beta \right)\;  
\label{current}, \eea where the sum over repeated Cartesian indices is understood, and $\Omega_{\alpha\beta}(\kk)$ is the many-body Berry curvature \[ \Omega_{\alpha\beta}(\kk)  = -2 \,\mbox{Im } \ev{\partial_{\kappa_\alpha} \Psi_{0\kk} | \partial_{\kappa_\beta} \Psi_{0\kk}} . \] 
The extensive quantity $\Omega_{\alpha\beta}(\kk) \dot{\kappa}_\beta$ is the many-electron anomalous velocity.  
In the static case ($\dot{\kk}=0$) no dc current may flow trough an insulating sample, ergo the ground-state energy $E_{0\kk}=E_0$ is $\kk$-independent; in metals, instead, $E_{0\kk}$ {\it does} depend on $\kk$ \cite{Kohn64}.

The linear conductivity is by definition \[ \sigma_{\alpha\beta}(\omega) = \frac{\partial j_\alpha(\omega)}{\partial \EE_\beta(\omega)} = \frac{\partial j_\alpha(\omega)}{\partial A_\beta(\omega)} \frac{d A(\omega)}{d \EE(\omega)} ; \label{r1} \] since $\EE(\omega) = i\omega A(\omega)/c$, causal inversion yields the last factor as \cite{Scalapino92} \[\frac{dA(\omega)}{d\EE(\omega)} = -\frac{ic}{\omega +i\eta} =  -c \left[ \pi \delta(\omega) + \frac{i}{\omega} \right] . \label{p2} \] At finite $\omega$, the linear response  $\partial j_\alpha(\omega)/ \partial A_\beta(\omega)$ is provided by time-dependent perturbation theory (i.e. Kubo formul\ae\ \cite{suppl}); here instead we only address the response to a dc macroscopic field. The physical perturbation is therefore static; it enters the Hamiltonian as a dynamical one in the adiabatic limit, owing to the vector-potential gauge, mandatory within PBCs \cite{nota2}.
Hence we set \[ \frac{\partial j_\alpha(\omega)}{\partial A_\beta(\omega)} \doteq \frac{\partial j_\alpha(0)}{\partial A_\beta(0)}, \label{adiab} \] where the symbol ``$\doteq$'' means ``equal in the dc limit''.

We chose the perturbing vector potential in the form ${\bf A}(t) = {\bf A}(\omega)\emi{\omega t}$, ergo we set \[ \kk(t) = \frac{e}{\hbar c} {\bf A}(\omega)\emi{\omega t} , \quad  \quad \dot{\kk}(t) = -\frac{i e \omega}{\hbar c} {\bf A}(\omega)\emi{\omega t} ,  \label{k2} \] whence (to lowest nonvanishing order in $\omega$):
\[ \kk \doteq \frac{e}{\hbar c} {\bf A}(0), \quad  \dot\kk \doteq  -\frac{i e \omega}{\hbar c} {\bf A}(0) . \label{nonvanishing} \]  
From \eqs{current}{nonvanishing} it follows  that \[ \frac{\partial j_\alpha(\omega)}{\partial A_\beta(\omega)} \doteq \frac{\partial j_\alpha(0)}{\partial A_\beta(0)} = -\frac{e^2}{\hbar c L^d} \left( \frac{1}{\hbar}\frac{\partial^2 E_0}{\partial {\kappa_\alpha} \partial {\kappa_\beta}} - i \omega\Omega_{\alpha\beta}(0) \right) . \label{p1} \]  The product of \equ{p1} times \equ{p2} yields the real parts of symmetric (longitudinal) and antisymmetric (transverse) dc conductivities as: \bea \mbox{Re } \sigma_{\alpha\beta}^{(+)}(\omega) &\doteq& \frac{\pi e^2}{\hbar^2 L^d} \frac{\partial^2 E_0}{\partial {\kappa_\alpha} \partial {\kappa_\beta}} \delta(\omega) = D_{\alpha\beta} \delta(\omega); \label{s1} \\ \mbox{Re } \sigma_{\alpha\beta}^{(-)}(\omega) &\doteq& \mbox{Re } \sigma_{\alpha\beta}^{(-)}(0) =- \frac{ e^2}{\hbar L^d} \Omega_{\alpha\beta}(0) . \label{s2} \eea Both these equations are not new, and can be alternatively obtained by the standard sum-over-states Kubo formul\ae\ \cite{suppl} in the $\omega \rightarrow 0$ limit. 

The present unconventional derivation has the virtue of being easily generalizable to nonlinear dc conductivity, which is the major focus of the present work. \equ{s1} is Kohn's milestone expression for the Drude term in longitudinal conductivity \cite{Kohn64}; the derivation given here is inspired by Ref. \cite{Scalapino92}.
As for \equ{s2}, it holds for either insulators or metals, for either $d=2$ or $d=3$, and yields the geometric (or intrinsic) term in the AHC; extrinsic effects are discussed in the final part of the present Letter. AHC is nonzero only if the Hamiltonian of \equ{kohn} breaks  T-symmetry at $\kk=0$ (see also the discussion below about symmetry).

The case of a two-dimensional insulator deserves a separate discussion. Transverse conductivity is quantized: \[ \sigma_{xy}^{(-)}(0) = - \frac{e^2}{h} C_1 ,\label{ch2} \] where $C_1\in {\mathbb Z}$ is a Chern number. This famous relationship was first established at the independent-electron level, where $C_1$ is also known as TKNN invariant \cite{Thouless82}; it was later generalized by Niu, Thouless, and Wu, who provided the many-body expression for $C_1$ \cite{Niu85}. Following Ref. \cite{Xiao10} (Sec. III.C) the same invariant is conveniently recast as \[ C_1 = \frac{1}{2\pi} \int_0^{\frac{2\pi}{L}} \!\! d\kappa_x \int_0^{\frac{2\pi}{L}} \!\! d\kappa_x  \;  {\sf \Omega}_{xy}(\kk) \label{chern} ; \] \equ{chern} is quantized because it is equivalent to the integral over a torus. In order to show this, we remind that in insulators the ground-state energy $E_{0\kk}$ is $\kk$-independent, and we observe that
whenever the components of $\kk - \kk'$ are integer multiples of $2\pi/L$, then the state $\ei{(\kk-\kk')\cdot \hat{\r}} \ket{\Psi_{0\kk}}$ is eigenstate of $\hat{H}_{\kk'}$ with the same eigenvalue as $\ket{\Psi_{0\kk}}$. The eigenstates which define ${\sf \Omega}_{xy}(\kk)$ have therefore the required toroidal periodicity: \[ \ket{\Psi_{0\kk'}} = \ei{(\kk-\kk')\cdot \hat{\r}} \ket{\Psi_{0\kk}} . \label{gauge} \] Since ${\sf \Omega}_{xy}(\kk)$ is gauge-invariant, an arbitrary $\kk$-dependent phase factor may relate the two members of \equ{gauge}. It is worth stressing that in the topological case a globally smooth periodic gauge does not exist; in other words we can enforce \equ{gauge} as it stands (with no extra phase factor) only locally, not globally; we also notice that \equ{gauge} may be regarded as the many-body analogue of the periodic gauge in band-structure theory \cite{Vanderbilt}.

\equ{chern} is independent of the $L$ value, and its integrand is extensive: therefore in the large-$L$ limit the integration domain contracts to a point: \[ C_1 = \frac{1}{2\pi} \left(\frac{2\pi}{L} \right)^2  \Omega_{xy}(0) .  \] By comparing this to \equ{s2} for $d=2$, \equ{ch2} is immediately retrieved.

Next we move on to deal with nonlinear conductivity; for the symmetric longitudinal term we adopt the same definitions as in Refs. \cite{Watanabe20,suppl}. The same logic as adopted above yields \[ \sigma_{\alpha\beta\gamma}^{(+)}(\omega_1,\omega_2) \doteq \frac{e^3}{\hbar^3 L^d} \frac{\partial^3 E_0}{\partial \kappa_\alpha \, \partial \kappa_\beta \, \partial \kappa_\gamma }\; \frac{i}{\omega_1 +i\eta} \frac{i}{\omega_2 +i\eta} :  \] not surprisingly, this is indeed identical to the recent finding of Ref. \cite{Watanabe20}.

In order to address the antisymmetric second-order term, we  expand the many-electron anomalous velocity  as \[ \Omega_{\alpha\beta}(\kk) \dot{\kappa}_\beta \simeq \Omega_{\alpha\beta}(0)\dot{\kappa}_\beta +  \partial_{\kappa_\gamma} \Omega_{\alpha\beta}(0) \; \dot{\kappa}_\beta {\kappa}_\gamma . \] The first term yields the AHC, \equ{s2}; we focus on the second term in the following, and we evaluate it in the adiabatic limit. \equ{current} yields, to second order, \bea j^{(2)}_\alpha(\omega) &\doteq& \frac{e}{L^d} \partial_{\kappa_\gamma} \Omega_{\alpha\beta}(0) \; \dot{\kappa}_\beta {\kappa}_\gamma  \nn &\doteq& - \frac{e^2}{\hbar L^d} \partial_{\kappa_\gamma} \Omega_{\alpha\beta}(0) \; \EE_\beta \, {\kappa}_\gamma  ,
\eea where the second equality owes to $\EEE(t) = - \hbar \dot\kk(t)/e$ in the dc limit; the $\kappa_\gamma$ factor is dealt with in the same way as in \eqs{p2}{k2}, i.e.  \[ \kappa_\gamma(t) = \frac{e}{\hbar c} A_\gamma(\omega) \emi{\omega t} = - \frac{i}{\omega + i \eta} \frac{e}{\hbar} \EE_\gamma(\omega)\emi{\omega t}. \] Therefore, to leading order in $\omega$, \bea j^{(2)}_\alpha(\omega) &\doteq& \frac{e^3}{\hbar^2 L^d} \partial_{\kappa_\gamma} \Omega_{\alpha\beta}(0) \frac{i}{\omega + i \eta}  \EE_\beta \, \EE_\gamma \label{major} \\ &\doteq& \frac{i}{\omega +i\eta} \chi_{\alpha\beta\gamma} \EE_\beta \EE_\gamma, \quad  \chi_{\alpha\beta\gamma} = \frac{ e^3}{\hbar^2 L^d}\partial_{\kappa_\gamma} \Omega_{\alpha\beta}(0) \nonumber .\eea
This is the major result of the present work: the sought for general NHC formula, which also applies to cases with interaction and/or disorder. For a crystalline system  of independent electrons, \equ{major} converges---in the large-sample limit---to the original Sodemann-Fu formula:
see \equ{sode} below.

 The real part of the $\omega$-dependent factor in \equ{major} equals $\pi \delta(\omega)$: the many-electron system undergoes a transverse free acceleration. One gets a dc current upon replacement of the infinitesimal $\eta$ with an inverse relaxation time $1/\tau$. This is in stark contrast with AHC, \equ{s2}, accounting for a $\tau$-independent dc current (some {\it extrinsic} AHC contributions are $\tau$-dependent; see below).
 
 As for the symmetry properties of \equ{major}, we remind that in presence of T-symmetry $\Omega_{\alpha\beta}(\kk) = - \Omega_{\alpha\beta}(-\kk)$, while in presence of I-symmetry $\Omega_{\alpha\beta}(\kk) = \Omega_{\alpha\beta}(-\kk)$ \cite{Xiao10}: therefore in a T-symmetric system $\Omega_{\alpha\beta}(0)=0$ and the AHC vanishes. In the case of NHC the parity is swapped: the gradient of $\Omega_{\alpha\beta}(\kk)$ is even in T-symmetric systems, and odd in I-symmetric systems. Therefore the NHC requires breaking of I-symmetry; in the special case of a T-symmetric and I-breaking system, nonzero transverse conductivity appears at second order only. 
 
 Since the responses to $\EE_\beta \EE_\gamma$ and to $\EE_\gamma \EE_\beta$ coincide, $\chi_{\alpha\beta\gamma}$ is symmetrical in the $\beta,\gamma$ indices, while instead it is antisymmetrical in the $\alpha,\beta$ and $\alpha,\gamma$ indices. Therefore the current is always orthogonal to the field: if---for an arbitrary $\EEE$ orientation---we set the $x$-axis along $\EEE$, then $j_x \propto \chi_{xxx}=0$, while $j_y\propto \chi_{yxx}$ and $j_z \propto \chi_{zxx}$ are not constrained to be zero by (this) symmetry.
  
At the independent-electron level (either Hartree-Fock or Kohn-Sham) the many-electron wavefunction is a Slater determinant of Bloch orbitals $\ket{\psi_{j\k}} = \ei{\k \cdot \r} \ket{u_{j\k}}$; we normalize them to one over the crystal cell. The Berry curvature of band $j$ is \cite{Vanderbilt} 
 \[ \tilde\Omega_{j,\alpha\beta}(\k) = -2\, \mbox{Im } \ev{\partial_{k_\alpha} u_{j\k} | \partial_{k_\beta} u_{j\k}} , \] and the many-body curvature per unit volume is \cite{suppl} \[ \frac{1}{L^d}\Omega_{\alpha\beta}(0) = \sum_j \int_{\rm BZ} \frac{d\k}{(2\pi)^d} f_j(\k) \, \tilde\Omega_{j,\alpha\beta}(\k) , \label{converg} \] where BZ is the Brillouin zone, and $f_j(\k)$ is the Fermi factor at $T=0$. 
 
The equality holds in the $L \rightarrow \infty$ limit. The convergence of \equ{converg} with $1/L$ has been indeed investigated by means of tight-binding simulations in the simple case of a Chern insulator, where the r.h.s. is quantized: Fig. 2 in Ref. \cite{rap135}.
 
It is then easy to prove \cite{suppl} that  \[ \frac{1}{L^d}\partial_{\kappa_\alpha} \Omega_{\alpha\beta}(0) = \sum_j \int_{\rm BZ} \frac{d\k}{(2\pi)^d} f_j(\k) \; \partial_{k_\alpha} \tilde\Omega_{j,\alpha\beta}(\k) , \] whence \[ \chi_{\alpha\beta\gamma} = \frac{e^3}{\hbar^2} \sum_j \int_{\rm BZ} \frac{d\k}{(2\pi)^d} f_j(\k) \; \partial_{k_\gamma} \tilde\Omega_{j,\alpha\beta}(\k) .  \label{sode} \] This is equivalent---in the single-band case---to the semiclassical expression which first appeared in the founding NHC paper by Sodemann and Fu \cite{Sodemann15}. The current induced by a monochromatic field of frequency $\omega$ has a dc (i.e. rectifying) component and a second-harmonic component. The adiabatic limit of the two terms is considered separately in Ref. \cite{Sodemann15}, hence a factor $1/2$ in each of them \cite{nota}.

In the final part of this Letter I revert  to AHC in order to comment on the extrinsic effects. 
First of all I stress the quite different role of the impurities between the AHC in metals and the quantized AHC in 2$d$ insulators: in the former case there must necessarily be extrinsic effects, while in the latter case extrinsic effects are ruled out. In fact---as a basic tenet of topology---any impurity has no effect on linear Hall conductivity insofar as the system remains insulating.

In a pristine metal the dc longitudinal conductivity is infinite: the Drude term is proportional to $\delta(\omega)$. Extrinsic mechanisms are necessary to warrant Ohm's law, and are accounted for by relaxation time(s) $\tau$; in absence of T-symmetry, extrinsic effects contribute to AHC as well. Two distinct mechanisms have been identified: they go under the name of ``side jump'' and ``skew scattering'' \cite{Nagaosa10}.  The side-jump term is nondissipative (independent of $\tau$). Since a crystal with impurities actually is a (very) dilute alloy, it was previously argued \cite{rap149} that the sum of the intrinsic and side-jump terms can be regarded as the intrinsic (geometrical) term of the dirty sample, whose AHC is given by \equ{s2} as it stands, provided that  the potential $\hat{V}$ includes the effect of the impurities. At the independent-electron level, the same effect can in principle be retrieved from the complementary real-space formulation of AHC \cite{rap153}. The other extrinsic term (skew scattering) is dissipative, proportional to $\tau$ in the single-relaxation-time approximation, and presumably cannot be explained by means of geometrical concepts. Remarkably, NHC is also proportional to $\tau$, yet it is a geometrical effect.

In this Letter I have started addressing linear dc conductivity (longitudinal and transverse), showing that their many-body expressions can be retrieved in an alternative way, making no use of the standard sum-over-states Kubo formul{\ae}; in this formulation AHC owes to the many-electron generalization of the anomalous velocity. Then I have adopted the same logic to second order in the field. In the longitudinal case the present approach retrieves the same result as in Ref. \cite{Watanabe20}; in the transverse case the quadratic expansion of the anomalous velocity yields a compact generalization of the semiclassical NHC formula of Ref. \cite{Sodemann15}. Even in presence of electron-electron interaction and/or disorder, NHC is dominated by the quantum geometry of the electronic system.
Finally, it is worth observing that---as it often happens when dealing with transport phenomena \cite{Xiao10,rap157}---the semiclassical NHC coincides with the exact one at the independent-electron level.

I thank Gabriele Bellomia and Ivo Souza for illuminating discussions, and for bringing some relevant papers to my attention. Work supported by the Office of Naval Research (USA) Grant No. N00014-17-1-2803.


\begin{thebibliography}{10}

\bibitem{Nagaosa10}
{ N. Nagaosa, J. Sinova, S. Onoda, A. H. MacDonald, and N. P. Ong, Rev. Mod.
  Phys. {\bf 82}, 1539 (2010)}.

\bibitem{Sodemann15}
{ I. Sodemann and L. Fu, Phys. Rev. Lett. {\bf 115}, 216806 (2015)}.

\bibitem{Matsyshyn19}
{ O. Matsyshyn and I. Sodemann, Phys. Rev. Lett. {\bf 123}, 246602 (2019)}.

\bibitem{Nandy19}
{ S. Nandy and I. Sodemann, Phys. Rev. B {\bf 100}, 195117 (2019)}.

\bibitem{Niu84}
{ Q. Niu and D. J. Thouless, J. Phys A {\bf 17}, 2453 (1984)}.

\bibitem{rap149}
{ R. Bianco, R. Resta, and I. Souza, Phys. Rev. B {\bf 90}, 125153 (2014)}.

\bibitem{Kohn64}
{ W. Kohn, Phys. Rev. {\bf 133}, {A171} (1964)}.

\bibitem{Xiao10}
{ D. Xiao, M.-C. Chang, and Q. Niu, Rev. Mod. Phys. {\bf 82}, 1959 (2010)}.

\bibitem{Scalapino92}
{ D. J. Scalapino, S. R. White, and S. C. Zhang, Phys. Rev. Lett. {\bf 18},
  2830 (1992)}.

\bibitem{suppl}
{ See Supplemental Material}.

\bibitem{nota2}
{ The scalar potential $\phi(\r) = - \EEE \cdot \r$ is incompatible with PBCs}.

\bibitem{Thouless82}
{ D. J. Thouless, M. Kohmoto, M. P. Nightingale, and M. den Nijs, Phys. Rev.
  Lett. {\bf 49}, 405 (1982)}.

\bibitem{Niu85}
{ Q. Niu, D. J. Thouless, and Y. S. Wu, Phys. Rev. B {\bf 31}, 3372 (1985)}.

\bibitem{Vanderbilt}
{ D. Vanderbilt, {\it Berry Phases in Electronic Structure Theory} (Cambridge
  University Press, Cambridge, 2018)}.

\bibitem{Watanabe20}
{ H. Watanabe and M. Oshikawa, Phys. Rev. B {\bf 102}, 165137 (2020)}.

\bibitem{rap135}
{ D. Ceresoli and R. Resta, Phys. Rev. B {\bf 76}, 012405 (2007)}.

\bibitem{nota}
{ The relationships between the vector and tensor forms of a Berry curvature
  are $\Omega_{\alpha\beta} = \varepsilon_{\alpha\beta\gamma} \Omega_\gamma$,
  $\Omega_\gamma = \frac{1}{2} \varepsilon_{\alpha\beta\gamma}
  \Omega_{\alpha\beta}$}.

\bibitem{rap153}
{ A. Marrazzo and R. Resta, Phys. Rev. B {\bf 95}, 121114(R) (2017)}.

\bibitem{rap157}
{ R. Resta, J. Phys. Condens. Matter {\bf 30}, 414001 (2018)}.

\end{thebibliography}

\end{document}


\title{Linear and nonlinear Hall conductivity in presence of interaction and disorder: Supplemental material}

\date{\today}



\maketitle 

\subsection*{Many-body Kubo formul\ae\ }

We define the matrix elements of the many-body velocity operator at $\kk=0$ :
 \bea {\cal R}_{n,\alpha\beta} &=& \mbox{Re }\langle \Psi_0 | \hat v_\alpha | \Psi_n \rangle \langle
\Psi_n | \hat v_\beta | \Psi_0 \rangle  , \\  {\cal I}_{n,\alpha\beta} &=& \mbox{Im }\langle \Psi_0 | \hat v_\alpha | \Psi_n \rangle \langle \Psi_n | \hat v_\beta | \Psi_0 \rangle ,
\label{vmat} \eea where ${\cal R}_{n,\alpha\beta}$ is symmetric and ${\cal I}_{n,\alpha\beta} $ antisymmetric; we further set $\omega_{0n} = (E_n - E_0)/\hbar$.
The longitudinal (symmetric) conductivity is: \[  \sigma_{\alpha\beta}^{(+)}(\omega) = D_{\alpha\beta} \left[ \delta(\omega) + \frac{i}{\pi \omega} \right] +\sigma_{\alpha\beta}^{(\rm regular)}(\omega) , \label{cond} \] 
\index{conductivity}
\[ D_{\alpha\beta} = \frac{\pi e^2}{ L^d} \left( \frac{N}{m} \delta_{\alpha\beta} - \frac{2}{\hbar} {\sum_{n\neq 0}}  \frac{{\cal R}_{n,\alpha\beta}  }{\omega_{0n}} \right) \label{drude} , \]
\bea \mbox{Re } \sigma_{\alpha\beta}^{(\rm regular)}(\omega) &=& \frac{\pi e^2}{\hbar L^d}  {\sum_{n\neq 0}} \frac{ {\cal R}_{n,\alpha\beta}}{\omega_{0n}} \delta(\omega - \omega_{0n}) , \; \omega>0 \label{s1} \\ \mbox{Im } \sigma_{\alpha\beta}^{(\rm regular)}(\omega) &=& \frac{2 e^2}{\hbar L^d}  {\sum_{n\neq 0}} \frac{ {\cal R}_{n,\alpha\beta}}{\omega_{0n}} \frac{\omega}{\omega_{0n}^2 - \omega^2} \label{s2} ; \eea  the Drude weight $D_{\alpha\beta}$ vanishes in insulators. 

The real part of longitudinal conductivity obeys the $f$-sum rule  \bea \int_0^\infty d \omega \; \mbox{Re } \sigma_{\alpha\beta} (\omega) &=& \frac{D_{\alpha\beta}}{2} + \int_0^\infty d \omega \; \mbox{Re } \sigma_{\alpha\beta}^{(\rm regular)} (\omega) \nn &=&\frac{\omega_{\rm p}^2}{8}\delta_{\alpha\beta} = \frac{\pi e^2 n}{2 m}\delta_{\alpha\beta} , \label{fsum} \eea where $n = N/L^d$ is the electron density and $\omega_{\rm p}$ is the plasma frequency. 

Using then the relationship \[ \ket{\partial_{\kappa_\alpha} \Psi_0} = - \sum_{n \neq 0} \ket{\Psi_n} \frac{\me{\Psi_n}{\hat v_\alpha }{\Psi_0}}{\omega_{0n}} , \label{der} \]
the Drude weight  can be recast as a geometrical property of the electronic ground state: \[  D_{\alpha\beta} = \frac{\pi e^2 N}{m L^d} \delta_{\alpha\beta} - \frac{2 \pi e^2}{\hbar^2 L^d} \mbox{Re } \me{\partial_{\kappa_\alpha} \Psi_0}{\,( \hat{H} - E_0 )\,}{\partial_{\kappa_\beta} \Psi_0} .  \] If we then start from the identity $ \me{\Psi_{0\kk}}{\,( \hat{H}_{\kk} - E_{0\kk} )\,}{\Psi_{0\kk}} \equiv 0 \label{iden} $, we take two derivatives, and we set  $\kk = 0$, we arrive at Kohn's famous expression for the Drude weight:  \[ D_{\alpha\beta} = \frac{\pi e^2}{\hbar^2 L^d} \frac{\partial^2 E_0}{\partial \kappa_\alpha \partial \kappa_\beta} , \] proved in the main text in an alternative way.

Transverse conductivity is nonzero only when T-symmetry is absent. The Kubo formul\ae\  for the transverse (antiymmetric) conductivity are:
\bea \mbox{Re } \sigma_{\alpha\beta}^{(-)}(\omega) &=& \frac{2e^2}{\hbar  L^d} {\sum_{n\neq 0}} \frac{{\cal I}_{n,\alpha\beta}}{\omega_{0n}^2 - \omega^2} \label{s3}
\\ \mbox{Im } \sigma_{\alpha\beta}^{(-)}(\omega) &=& \frac{\pi e^2}{\hbar L^d} {\sum_{n
\neq 0}} \frac{{\cal I}_{n,\alpha\beta}}{\omega_{0n}} \delta(\omega - \omega_{0n}) , \; \omega> 0 . \label{s4} \eea 
Using again \equ{der} the dc transverse conductivity is easily recast in terms of the many-body Berry curvature at $\kk=0$: \[ \mbox{Re } \sigma_{\alpha\beta}^{(-)}(0) = - \frac{e^2}{\hbar  L^d}  \Omega_{\alpha\beta}(0) ; \label{main} \] the expression holds for metals and insulators, in either 2$d$ or 3$d$. Even this expression in the main text is obtained in an alternative way, by means of the anomalous velocity in its many-body formulation.

\subsection{Second-order conductivity}

In the time domain, the second-order conductivity is by definition \[ \sigma_{\alpha\beta\gamma} = \frac{\delta j_\alpha(t)}{\delta \EE_\beta(t') \, \delta \EE_\gamma(t'')} , \] or equivalently (for a time-independent system) \[ j_\alpha(t) = \frac{1}{2} \int_0^\infty \!\!\!\! dt' \int_0^\infty \!\!\!\! dt'' \; \sigma_{\alpha\beta\gamma}(t - \!\! t', t \!\!\ - \!\! t'') \EE_\beta(t') \EE_\gamma(t'') \label{sigma2} ; \] therein all quantities are real and $\sigma$ is causal.

The convolution theorem yields \[ \frac{\partial^2 j_\alpha(\omega_1+\omega_2)}{\partial \EE_\beta(\omega_1) \partial \EE_\gamma(\omega_2)}  = \int_0^\infty  \!\!\!\! dt' \int_0^\infty \!\!\!\! dt'' \; \ei{\omega_1 t'} \ei{\omega_2 t''}\sigma_{\alpha\beta\gamma}(t',t'') , \label{convolution} \] where all quantities are complex. The second-order conductivity $\sigma(\omega_1,\omega_2)$ is identified with \equ{convolution} in Ref. \cite{Watanabe20}, and consistently in our Eq. (17) in the main text.

The current induced to second order by a monochromatic field $\EEE(\omega)$ at finite frequency has a dc (i.e. rectifying) component, and a second-harmonic component at frequency $2\omega$; in the longitudinal case the dc response goes under the name of ``shift photocurrent''.
Since the real part of a product is obviously not equal to the product of the real parts, the second-order conductivity $\sigma(\omega,\pm \omega)$ can be defined in different ways, related to \equ{convolution} by a factor of two \cite{Young12} or even four \cite{Ibanez18}.

In the main text we deal with the transverse quadratic response, in the $\omega \rightarrow 0$ limit only. We avoid mentioning $\sigma(\omega,\pm \omega)$ altogether, and we explicitly express the current in terms of the field. It is worth reminding that fields can be dealt with directly at the semiclassical level, but the scalar potential $\phi(\r) = - \EEE \cdot \r$ is incompatible with PBCs.
Therefore we deal here with time-dependent vector potentials in the adiabatic limit. 

\subsection*{Independent-electron results}

The independent-electron AHC of a pristine crystal is \[  \mbox{Re } \sigma_{\alpha\beta}^{(-)}(0) = - \frac{e^2}{\hbar} \sum_j \int_{\rm BZ} \frac{d\k}{(2\pi)^d} f_j(\k) \, \tilde\Omega_{j,\alpha\beta}(\k) , \] where BZ is the Brillouin zone, $f_j(\k)$ is the Fermi factor at $T=0$, and $\tilde\Omega_{j,\alpha\beta}(\k)$ is the Berry curvature of band $j$; comparison to \equ{main} yields: \[ \frac{1}{L^d} \Omega_{\alpha\beta}(0) = \sum_j \int_{\rm BZ} \frac{d\k}{(2\pi)^d} f_j(\k) \, \tilde\Omega_{j,\alpha\beta}(\k) . \] The $L \rightarrow \infty$ is implicitly understood in the l.h.s.; it is instead explicit in the r.h.s., given that the Bloch vector therein is a continuous variable.

When $\kk \neq 0$ is set in Kohn's Hamiltonian $\hat{H}_{\kk}$, the corresponding Kohn-Sham periodic orbitals $\ket{u_{j\k}}$ are eigenstates of the single-particle Hamiltonian \[ \emi{\k \cdot \r}H_{\kk}\ei{\k \cdot \r} = 
\frac{1}{2m} 
\left[ \p + \frac{e}{c}{\bf A}(\r) + \hbar \k + \hbar \kk \right]^2 \!\!+ V_{\rm KS} , \] where $V_{\rm KS}$ is the Kohn-Sham potential, hence \[ \frac{1}{L^d}  \Omega_{\alpha\beta}(\kk) = \sum_j \int_{\rm BZ} \frac{d\k}{(2\pi)^d} f_j(\k) \, \tilde\Omega_{j,\alpha\beta}(\k+ \kk) . \] \[ \frac{1}{L^d}\partial_{\kappa_\alpha} \Omega_{\alpha\beta}(0) = \sum_j \int_{\rm BZ} \frac{d\k}{(2\pi)^d} f_j(\k) \; \partial_{k_\alpha} \tilde\Omega_{j,\alpha\beta}(\k) . \] 
